\documentclass[aps,prd,twocolumn,groupedaddress,nofootinbib,showpacs,eqsecnum,amsmath,amssymb,useAms]{revtex4}

\usepackage{euscript,graphicx,amssymb,subfigure}

\usepackage{epsfig,graphics}
\usepackage{amsthm,dsfont}
\usepackage{amsfonts}
\usepackage{amsmath}
\usepackage{amssymb}
\usepackage{euscript}
\usepackage{color}
\usepackage{subfigure}
\usepackage{fontenc}
\usepackage{textcomp}
\usepackage{relsize}

\def\lsim{\mathrel{\rlap{\lower3.5pt\hbox{\hskip0.5pt$\sim$}}\raise0.5pt\hbox{$<$}}}
\def\gsim{~\rlap{$>$}{\lower 1.0ex\hbox{$\sim$}}}

\begin{document}

\title{Gravitational wave energy spectrum of hyperbolic encounters}

\author{Lorenzo De Vittori$^1$\footnote{\tt lorenzo@physik.uzh.ch}, Philippe Jetzer$^1$\footnote{\tt jetzer@physik.uzh.ch} and Antoine Klein$^{1,2}$\footnote{{\tt aklein@physics.montana.edu}}}

\affiliation{$^1$ Universit$\ddot{a}$t Z$\ddot{u}$rich, Institut f$\ddot{u}$r Theoretische Physik, 8057 Z$\ddot{u}$rich, Switzerland\\
$^2$ Montana State University, Departement of Physics, Bozeman MT 59717, USA}

\begin{abstract}
The emission of gravitational waves is studied for a system of massive objects interacting on hyperbolic orbits
within the quadrupole approximation following the work of Capozziello et al. \cite{Capozziello}. Here we focus 
on the derivation of an analytic formula for the energy spectrum of the emitted waves.
We checked numerically that our formula is in agreement with the two limiting cases
for which results were already available: for  
the eccentricity $\varepsilon=1$, the parabolic case whose 
spectrum was computed by Berry and Gair \cite{BerryGair}, and the large $\varepsilon$ limit
with the formula given by Turner \cite{Turner}. 
\end{abstract}


\maketitle

\section{Introduction}

Einstein predicted already in 1916 that accelerated masses should emit gravitational waves.
The detection of such kind of waves would open a new window in the exploration of our universe.
In the last years technology has improved very rapidly, and it is now believed that the precision we reached should  
enable the direct detection of gravitational waves in few years, both with ground based and space based detectors
such as e.g. the proposed eLISA mission.
It is, therefore, interesting to study the dynamics of typical systems and their emission of gravitational waves
and in particular their frequency spectrum, in order to know at which wave-length range we should expect 
gravitational radiation.\\

For the cases of binary systems or spinning black holes on circular and elliptical orbits 
the resulting energy spectra have already been well studied 
\cite{peters1,peters2}. The energy spectrum for parabolic encounters has been computed either
by direct integration along unbound orbits \cite{Turner} or more recently by taking the limit
of the Peters and Mathews energy spectrum for eccentric Keplerian binaries \cite{BerryGair}.

The emission of gravitational waves from a system of massive objects interacting on hyperbolic
trajectories using the quadrupole approximation has been studied by Capozziello et al. \cite{Capozziello}
and analytic expressions for the total energy output derived.
However, the energy spectrum has been computed only for the large eccentricity ($\varepsilon \gg 1$)
limit \cite{Turner}.
In this paper we derive the energy spectrum for hyperbolic encounters for all values $\varepsilon \geq 1$
and we give an analytic expression for it in terms of Hankel functions. We checked numerically 
that our result in the limit of $\varepsilon=1$ is in agreement with the one for parabolic encounters
\cite{BerryGair} and for large eccentricities with the 
result given in \cite{Turner}.

\section{Theoretical framework}\label{sec:theory}

Gravitational waves (GWs) are solutions of the linearized field equations of General Relativity and the radiated 
power to leading order is given by Einstein's quadrupole formula, as follows
\begin{equation}\label{eq:poweremission}
 P = \frac{G}{45 \, c^5} \langle \dddot{D}_{ij} \, \dddot{D}_{ij} \rangle~,
\end{equation}
where we used as definition for the second moment tensors $M_{ij} := \frac{1}{c^2} \int T^{00} x_i x_j \, \mathrm{d}^3x$, and
for the quadrupole moment tensor $D_{ij} := 3 M_{ij} - \delta_{ij} M_{kk}$. Here and in the following dots denote time derivatives
\footnote{Note that often in the literature (e.g. \cite{Maggiore}) we also find the notation $Q_{ij}:=\frac{D_{ij}}{3}=M_{ij} - \frac{1}{3}\delta_{ij} M_{kk}$,
equation (\ref{eq:poweremission}) reads then: $P_{\textmd{quad}} = \frac{G}{5 \, c^5} \langle \,\dddot{Q}_{ij} \, \dddot{Q}_{ij} \rangle$.
Here and in the following we will use the notation given by \cite{Capozziello} and \cite{LL}.}.

The quantity $M_{ij}$ depends on the trajectories of the involved masses, and can easily be computed for all type of Keplerian trajectories.
To compute the power spectrum, i.e. the amplitude of radiated power per unit frequency,
requires a Fourier transform of equation (\ref{eq:poweremission}), which is rather involved (for the elliptical case see e.g. \cite{Maggiore}),
and we will derive it below for hyperbolic encounters.

In Fig. 1 the geometry of an hyperbolic encounter is represented with the most important quantities we will use.
Since we will compare our results with those of \cite{BerryGair} and \cite{Capozziello}, it is important to note that not all these quantities are independent
from each other, and we will need to know the relations between them.
Notice that we assume that the gravitational energy loss during the encounter is negligible
and thus that the Keplerian hyperbolic trajectory is a good approximation of the orbit.
Clearly, this assumption does depend on the mass ratio and on the distance of closest approach. In the considered cases, where the masses are similar, this holds very well.

The eccentricity $\varepsilon$ of the hyperbola is (see e.g. \cite{Maggiore})
\begin{equation} \label{eq:eccentricity}
 \varepsilon := \sqrt{1+\frac{2\,E\,L^2}{\mu\,\alpha^2}}~,
\end{equation}
where $E = \frac{1}{2}\mu\,v_0^2$ ($E$ is a conserved quantity for which we can take the energy at $t=-\infty$),
$v_0$ being the velocity of the incoming mass $m_1$ at infinity, the angular momentum
$L=\mu\,b\,v_0$, the impact parameter $b$, the reduced mass $\mu := \frac{m_1\,m_2}{m_1+m_2}$, the total mass $m:=m_1+m_2$, and the parameter $\alpha := G\,m\,\mu$.\\

We notice that the orbit is characterized by only four quantities given as initial conditions: $v_0$, $b$ and $m_{1,2}$.
All the other parameters can be expressed as functions of this fundamental set.
We can for instance rewrite the eccentricity as $\varepsilon = \sqrt{1+v_0^4\,b^2\,/\,G^2\,m^2} = \varepsilon \, (v_0,b,m)$,
or the semi-major axis $a=\alpha/\mu v_0^2$, the angle at periastron through $\cos\varphi_0 = -1/\varepsilon$, or the radius at periastron as $r_{min} = \frac{G m}{v_0^2}(\varepsilon-1)$, and so on.

\begin{figure}[!htb]
 \begin{center}
  \includegraphics[width=0.47\textwidth]{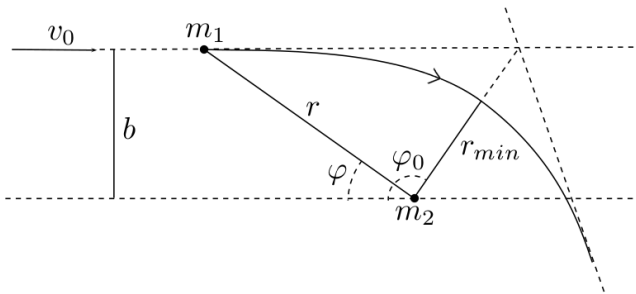}
  \label{fig:geom_of_the_encounter}
  \caption{The geometry of an hyperbolic encounter and its most important parameters.}
 \end{center}
\end{figure}
Setting the angle of the incident body to $\varphi=0$ at initial time $t=-\infty$, the radius of the trajectory as a function of the angle and as a function of time is given by
\begin{equation*}
 r(\varphi)=\frac{a\,(\varepsilon^2-1)}{1+\varepsilon \, \cos (\varphi-\varphi_0)} ~,\quad r(\xi)=a\,(\varepsilon\,\cosh\xi-1) 
\end{equation*}
with the time parametrized by $\xi$ through the relation $t(\xi)=\sqrt{\frac{\mu\,a^3}{\alpha}}\,(\varepsilon\,\sinh\xi-\xi)$,
where $\xi$ goes from $-\infty$ to $+\infty$.\\
Expressing this in Cartesian coordinates in the orbital plane, we finally get the equations for hyperbolic trajectories
\begin{subequations}
\begin{align}
 x(\xi)&=a\,(\varepsilon-\cosh\xi)~,\\
 y(\xi)&=a\,\sqrt{\varepsilon^2-1}\,\sinh\xi~.
\end{align}
\end{subequations}

\section{Power spectrum of Gravitational waves from hyperbolic paths}\label{sec:powerspectrum}

\subsection{Power emitted per unit angle}\label{subsec:powerunitangle}

In \cite{Capozziello} the computation of the power emitted as a function of the angle, as well as the total energy emitted by the system has been already carried out.
Here, we briefly present these computations, whose results we will then use.

First we compute the energy and angular momentum of a body within a gravitational potential $\Phi(r)$. In the plane of the orbit
the velocity can be written in terms of a tangent and a perpendicular component
\begin{equation*}
 \textbf{v} = v_r \, \hat{\bf{r}} + v_{\varphi} \, \hat{\boldsymbol{\varphi}} \textmd{, \quad where \;}
 v_r = \frac{dr}{dt}, \quad v_{\varphi} = r\,\frac{d\varphi}{dt}~,
\end{equation*}
where vectors are represented by bold symbols.
Thus the total energy per unit mass of the system and the angular momentum can be written as
\begin{align*}
 E := \frac{1}{2}\,v^2+\Phi(r) =& \frac{1}{2} \, \bigg( \frac{dr}{dt} \bigg)^2 + \frac{1}{2} \, r^2 \,\bigg(\frac{d\varphi}{dt} \bigg)^2 + \Phi(r)~,\\
 L := \textbf{r} \times \textbf{v} =& r^2 \,\frac{d\varphi}{dt}~.
\end{align*}
Putting these equations together, using the substitution $u:=1/r$ with $r^2 = L/\dot{\varphi}$ and rearranging, we get
\begin{equation*}
 \frac{2\,E}{L^2} = \frac{\dot{u}^2}{\dot{\varphi}^2} + u^2 + \frac{2\,\Phi}{L^2} = \bigg(\frac{du}{d\varphi}\bigg)^2 + u^2 + \frac{2\,\Phi}{L^2}~.
\end{equation*}
Since $E$ and $L$ are conserved quantities, the derivative of the last expression with respect to $u$ gives:
\begin{equation*}
 0 = \frac{d^2u}{d\varphi^2} + u + \frac{1}{L^2} \, \frac{d\Phi}{du} \quad \Leftrightarrow \quad \frac{d^2u}{d\varphi^2} + u = \frac{G\,m_2}{L^2}~.
\end{equation*}
This is an inhomogeneous linear differential equation of second order, which has the following solution
\begin{equation*}
 u(\varphi) = B\,\cos(\varphi-\varphi_0) + \frac{G\,m_2}{L^2}~,
\end{equation*}
and substituting back to r we have $\dot{r} = B\,L\,\sin(\varphi-\varphi_0)$.
$B$ is a constant depending on the initial conditions and $\varphi_0$ is the polar angle corresponding to the periastron
distance, i.e. the distance of closest approach between the two interacting bodies (see Fig. 1). Its relation to the eccentricity is given by $\varepsilon = -1/\cos\varphi_0$.

Using the initial condition for the velocity, and the standard procedure to reduce the two-body problem to a single reduced mass particle
moving in a gravitational field generated by the total mass, we see that the orbit of this reduced mass particle reads
\begin{equation}
 r(\varphi) = \frac{b \,  \sin \varphi_0}{\cos(\varphi-\varphi_0)-\cos \varphi_0}~.
\end{equation}

In order to compute the emitted power $P$, given by the quadrupole formula, as a function of the angle, it is convenient to 
rewrite the Cartesian coordinates $x_i$ in spherical coordinates (the plane of the orbit corresponds to $\vartheta = \frac{\pi}{2}$)
\footnote{Note that $P$ depends on $D_{ij}$, which also depends on the $x_i$ present in the integral definition of $M_{ij}$.}
\begin{align*}
 x &= r \cos\varphi \sin\vartheta = r \cos\varphi~, \\
 y &= r \sin\varphi \sin\vartheta = r \sin\varphi~, \\
 z &= r \cos\vartheta = 0~.
\end{align*}
For the second momenta tensors $M_{ij}$ we get accordingly
\begin{equation}\label{eq:M_ij}
 \begin{split}
 M_{11} &= \mu x^2 = \mu r^2 \cos^2\varphi~,\\
 M_{22} &= \mu y^2 = \mu r^2 \sin^2\varphi~, \\
 M_{12} &= \mu x y = \mu r^2 \cos\varphi \sin\varphi = M_{21}~,\\
 M_{32} &= M_{23} = M_{33} = M_{13} = M_{31} = 0~.
 \end{split}
\end{equation}
The term $\dddot{D}_{ij} \dddot{D}_{ij} = \sum_{i,j} \dddot{D}_{ij} \dddot{D}_{ij}$ in the expression for the radiated power can now be simplified to give
\begin{equation*}\label{eq:quadrupole}
 \textlangle \dddot{D}_{ij} \dddot{D}_{ij} \textrangle \; = 6 \; \textlangle \dddot{M}_{11}^2 + \dddot{M}_{22}^2 + 3 \dddot{M}_{12}^2 - \dddot{M}_{11} \dddot{M}_{22} \textrangle~.
\end{equation*}
In order to compute this value explicitly, keeping $\varphi$ as a variable instead of $t$, we have to transform derivatives in time in derivatives in $\varphi$ and $r$.
This yields
\begin{equation} \label{eq:P_capozziello}
 P(\varphi) = - \frac{32 \,G\,L^6 \mu^2}{45\,c^5\,b^8} f(\varphi,\varphi_0)~,
\end{equation}
with
\begin{equation*}
 \begin{split}
  f(\varphi,\varphi_0) := \frac{\sin(\varphi_0 - \frac{\varphi}{2})^4 \; \sin(\frac{\varphi}{2})^4 }{\tan(\varphi_0)^2 \;
  \sin(\varphi_0)^6} \cdot \bigg(150 + 72 \cos(2 \varphi_0) + \\ + \; 66 \cos(2 (\varphi_0 - \varphi)) - 144 \; (\cos(2 \varphi_0 - \varphi) - \cos(\varphi))\bigg)~,
 \end{split}
\end{equation*}
which is the result obtained in \cite{Capozziello}.
In Fig. 2 we plot the radiated power $P$ as a function of the angle. 
\begin{figure}[!htb]
 \begin{center}
  \includegraphics[width=0.47\textwidth]{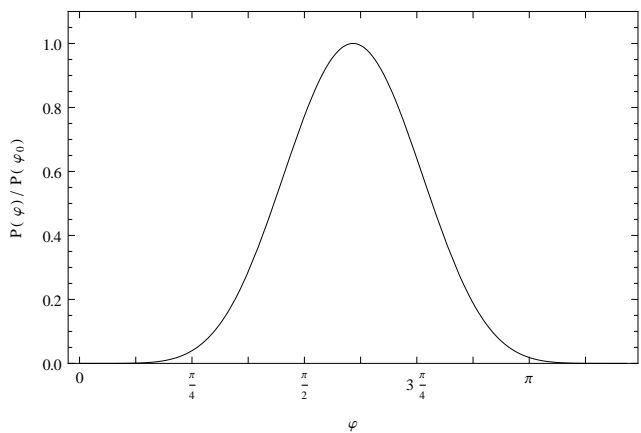}
  \label{fig:P_phi}
  \caption{Radiated power as a function of the angle during an hyperbolic encounter, for $\varepsilon=3$,
  i.e. $\varphi_0 \simeq 0.6 \pi$, according to the relation $\varepsilon=-1/\cos\varphi_0$.}
 \end{center}
\end{figure}\\
Eq.  (\ref{eq:P_capozziello}) can also be written as follows, using the Schwarzschild radius $r_s:=2Gm/c^2$, and $L=v_0\,b$
\begin{equation*}
 P = \frac{dE}{dt} = - \frac{4 \, r_s v_0^6\,\mu}{45\,c^3\,b^2} f(\varphi,\varphi_0)~.
\end{equation*}
Setting $t=0$ at periastron, the total energy radiated in GWs by the system during the interaction is given by
\begin{equation}
 \Delta E = \int\limits_{-\infty}^\infty \!\bigg|\frac{dE}{dt}\bigg|\,\mathrm{d}t~.
\end{equation}
Since we know $\frac{dE}{dt}$ as function of $\varphi$ rather than $t$, we perform a variable change in the integration
and get
\begin{equation*}
 \Delta E = \frac{4 \, r_s \,v_0^5\,\mu}{45\,c^3\,b}\,\int_{\varphi_1}^{\varphi_2} \! \frac{\sin^2\varphi_0 \, f(\varphi,\varphi_0)}{[\cos(\varphi-\varphi_0)-\cos\varphi]^2} \,\mathrm{d}\varphi~,
\end{equation*}
which can be evaluated taking $\varphi_1 = 0$ as initial angle, and $\varphi_2 = 2 \varphi_0$ as the final angle
\begin{equation}\label{eq:Delta_E}
 \Delta E = \frac{32\,G\,\mu^2\,v_0^5}{b\,c^5}\,F(\varphi_0)~,
\end{equation}
with
\begin{equation*}
 \begin{split}
  F(\varphi_0) =& \frac{1}{720\,\tan^2\varphi_0 \, \sin^4 \varphi_0}\times[2628 \varphi_0 + 2328 \varphi_0 \cos 2\varphi_0\\
  +&144 \varphi_0 \cos4\varphi_0 - 1948\sin 2\varphi_0 - 301 \sin 4 \varphi_0]~.
 \end{split}
\end{equation*}

This means that the total radiated energy of the system can be determined knowing 
the parameters $b$ and $v_0$, and of course the mass $\mu$.

In deriving the above equation the implicit assumption has been made, that the energy loss 
doesn't change the path of the body in the gravitational field, which is reasonable 
since the emitted energy in form of gravitational waves is rather small.
As stated above, this holds very well for the cases we consider here, where the mass ratio is about $1$.

\subsection{Power spectrum}\label{subsec:powerspec}

We compute now $P(\omega)$, the Fourier transform of $P(t)$,
which describes the distribution of the amplitude of the power emitted in form of gravitational waves depending on the frequency.
In the following we use the convention
\begin{equation*}
 \hat{f}(\omega):=\int_{-\infty}^{\infty} \! f(t)\;e^{-i\omega t}\;\mathrm{d}t
\end{equation*}
In \cite{LL} and \cite{Longair} some hints are given when solving the analogous problem in electrodynamics.
The crucial idea is to use Parseval's theorem on the integration of Fourier transforms first,
and then to express some quantities in terms of Hankel functions.
This allows to compute in an easier way the Fourier transform of $P(t)$, for which we use the expression given in 
eq. (\ref{eq:poweremission})
\begin{equation*}
 \begin{split}
 &\Delta E = \!\int \! P(t)\mathrm{d}t = \!\int \! P(\omega)\mathrm{d}\omega =
 \gamma \!\int \! \textlangle \, \dddot{D}_{ij}(t) \, \dddot{D}_{ij}(t) \, \textrangle\,\mathrm{d}t=\\
 &\gamma \int \! (|\widehat{\dddot{D}_{11}}(\omega)|^2 + |\widehat{\dddot{D}_{22}}(\omega)|^2
 + 2 |\widehat{\dddot{D}_{12}}(\omega)|^2 + |\widehat{\dddot{D}_{33}}(\omega)|^2 )\,\mathrm{d}\omega~,
 \end{split}
\end{equation*}
where $\widehat{\dddot{D}_{ij}}(\omega)$ represents the Fourier transform of $\dddot{D}_{ij}(t)$, and we introduced the constant $\gamma := -\frac{G}{45c^5}$.\\

It is easy to see that the last equation represents the total amount of energy dissipated in the encounter.
Therefore, the integrand in the last line has to be equal to the power dissipated per unit frequency $P(\omega)$, i.e. :
\begin{equation}\label{eq:P_omega}
 P(\omega)= \gamma \, \big( |\widehat{\dddot{D}_{11}}(\omega)|^2 + |\widehat{\dddot{D}_{22}}(\omega)|^2 + 2 |\widehat{\dddot{D}_{12}}(\omega)|^2 + |\widehat{\dddot{D}_{33}}(\omega)|^2 \big)~.
\end{equation}

As next, we need to compute the $\widehat{\dddot{D}_{ij}}(\omega)$, to square their norm and sum them together in order to get the power spectrum.
Note that first we compute the Fourier transform of $D_{ij}(t)$, then we differentiate three times, and only at the end we take the square of their norm.
Computing the $D_{ij}$ explicitly - keeping in mind that we use the time parametrization $t(\xi) = \sqrt{\mu\,a^3/\alpha}\,(\varepsilon\,\sinh\xi-\xi)$ - we get:
\begin{equation}
 \begin{split}
 D_{11}(t) &= \frac{a^2 m}{2} \, ((3-\varepsilon^2) \cosh 2\xi - 8\,\varepsilon \cosh \xi)~,\\
 D_{22}(t) &= \frac{a^2 m}{2} \, (4\,\varepsilon \cosh \xi + (2\,\varepsilon^2-3) \cosh 2\xi)~,\\
 D_{33}(t) &= \frac{-a^2 m}{2} \, (4\,\varepsilon \cosh \xi + \varepsilon^2 \cosh 2\xi)~,\\
 D_{12}(t) &= \frac{3 \, m \, a^2}{2} \,\sqrt{\varepsilon^2-1} \,(2 \, \varepsilon \sinh \xi - \sinh 2\xi )~.
 \end{split}
\end{equation}
The Fourier transform of the third derivatives of $D_{ij}(t)$ is given by 
\begin{equation}\label{eq:3derivative}
\widehat{\dddot{D}_{ij}}(\omega) = i \omega^3 \, \widehat{D_{ij}}(\omega)~,
\end{equation}
thus we have just to compute $\widehat{D_{ij}}(\omega)$.
To compute the Fourier transforms we can closely follow
the calculations performed e.g. in \cite{LL}, where the
similar problem in electrodynamics of the emitted power spectrum for
scattering charged particles on hyperbolic orbits is treated.
In particular the following Fourier transforms are used
\begin{equation} \label{eq:HankelRelations}
  \widehat{\sinh \xi} = - \frac{\pi}{\omega \varepsilon} H_{i \nu}^{(1)}(i \nu \varepsilon) \; , \; \widehat{\cosh \xi} = - \frac{\pi}{\omega} H_{i \nu}^{(1)}\textquotesingle(i \nu \varepsilon)~,
\end{equation}
\begin{equation} \label{eq:HankelDerivative}
 H_{\tilde{\alpha}}^{(1)}\textquotesingle(x) = \frac{1}{2} (H_{\tilde{\alpha}-1}^{(1)}(x)-H_{\tilde{\alpha}+1}^{(1)}(x))~,
\end{equation}
where $H_{\tilde{\alpha}}^{(1)}(x)$ is the Hankel function of the first kind of order $\tilde{\alpha}$, defined as $J_{\tilde{\alpha}}(x) + i Y_{\tilde{\alpha}}(x)$,
with $J_{\tilde{\alpha}}(x), Y_{\tilde{\alpha}}(x)$ the Bessel functions of first and second kind, respectively; and where the 
frequency $\nu$ is defined as $\nu := \omega \sqrt{\frac{\mu a^3}{\alpha}}$
(the parameters $\mu$, $a$ and $\alpha$ as defined previously).

\noindent Taking the above results for $D_{ij}(t)$ we get\footnote{Notice that we obtain also constant terms in these four equations. However, they can be dropped, since we can freely change
the origin of coordinates while keeping invariant the description of the quadrupole radiation, as explained e.g. in \cite{Maggiore} in section 3.3.5.
Note also that in fact these terms would in any case vanish for merely mathematical reasons, since the Fourier transform would multiply
them with a Dirac delta function $\delta(\omega)$ and a factor $(-i\omega)^3$ from the third derivative.
This expression vanishes for all $\omega \neq 0$ because of the $\delta(\omega)$, and for $\omega = 0$ because of the multiplying factor $(-i\omega)^3$.}
\begin{equation*}
\begin{split}
 \widehat{D_{11}}(\omega) &= \frac{a^2\,m\,\pi}{4\,\omega}\,[16\,\varepsilon\,H_{i \nu}^{(1)}\textquotesingle(i\nu\varepsilon)+(\varepsilon^2-3)\,H_{i\nu}^{(1)}\textquotesingle(i\nu\varepsilon/2)]~,\\
 \widehat{D_{22}}(\omega) &= \frac{a^2\,m\,\pi}{4\,\omega}\,[(3-2\,\varepsilon^2)\,H_{i \nu}^{(1)}\textquotesingle(i\nu\varepsilon/2)-8\,\varepsilon\,H_{i\nu}^{(1)}\textquotesingle(i\nu\varepsilon)]~,\\
 \widehat{D_{33}}(\omega) &= \frac{a^2\,m\,\pi}{4\,\omega}\,[8\,\varepsilon\,H_{i \nu}^{(1)}\textquotesingle(i\nu\varepsilon)+\varepsilon^2\,H_{i\nu}^{(1)}\textquotesingle(i\nu\varepsilon/2)]~,\\
 \widehat{D_{12}}(\omega) &= \frac{3\,a^2\,m\,\pi}{4\,\omega\,\varepsilon}\,\sqrt{\varepsilon^2-1}\,[H_{i \nu}^{(1)}(i\nu\varepsilon/2)-4\,\varepsilon\,H_{i\nu}^{(1)}(i\nu\varepsilon)]~.
\end{split}
\end{equation*}
Inserting this result into eq. (\ref{eq:P_omega}), using eq. (\ref{eq:3derivative}), we get the power spectrum of the gravitational wave emission for hyperbolic encounters
\vspace{0.1in}
\begin{equation} \label{eq:P_omega_mio}
 P(\omega)=- \frac{G\,a^4\,m^2\,\pi^2}{720\,c^5} \; \omega^4 \; F_{\varepsilon}(\omega)~,\vspace{0.1in}
\end{equation}
where the function $F_{\varepsilon}(\omega)$ is
\begin{align*}
 \begin{split}
 F_{\varepsilon}(\omega) = &|[16\,\varepsilon\,H_{i \nu}^{(1)}\textquotesingle(i\nu\varepsilon)+(\varepsilon^2-3)\,H_{i\nu}^{(1)}\textquotesingle(i\nu\varepsilon/2)]|\,^2 \,+\\
 &|[(3-2\,\varepsilon^2)\,H_{i \nu}^{(1)}\textquotesingle(i\nu\varepsilon/2)-8\,\varepsilon\,H_{i\nu}^{(1)}\textquotesingle(i\nu\varepsilon)]|\,^2 \,+\\
 &|[8\,\varepsilon\,H_{i \nu}^{(1)}\textquotesingle(i\nu\varepsilon)+\varepsilon^2\,H_{i\nu}^{(1)}\textquotesingle(i\nu\varepsilon/2)]|\,^2 \,+\\
 &\frac{9\,(\varepsilon^2-1)}{\varepsilon^2}\,|[H_{i \nu}^{(1)}(i\nu\varepsilon/2)-4\,\varepsilon\,H_{i\nu}^{(1)}(i\nu\varepsilon)]|\,^2~.\\
 \end{split}
\end{align*}

In Fig. 3 the function $\omega^4 \, F_{\varepsilon}(\omega)$ is plotted for some some values of $\varepsilon$:
this is the frequency power spectrum of gravitational radiation emitted by an hyperbolic encounter.
Unfortunately the expression for $F_{\varepsilon}(\omega)$ is rather complicated and we could not find an analytical way to simplify it.
We thus made some numerical tests to check its validity and clearly the integral of (\ref{eq:P_omega_mio}) has to be equal to $\Delta E$ in (\ref{eq:Delta_E}),
which was obtained by integrating over the power emitted per unit frequency, i.e.:
\begin{equation}
 \int_0^{\infty} \!P(\omega) \,\mathrm{d}\omega = \Delta E~.\\
\end{equation}

\begin{figure}[!htb]\vspace{0.08in}
 \begin{center}
  \includegraphics[width=0.47\textwidth]{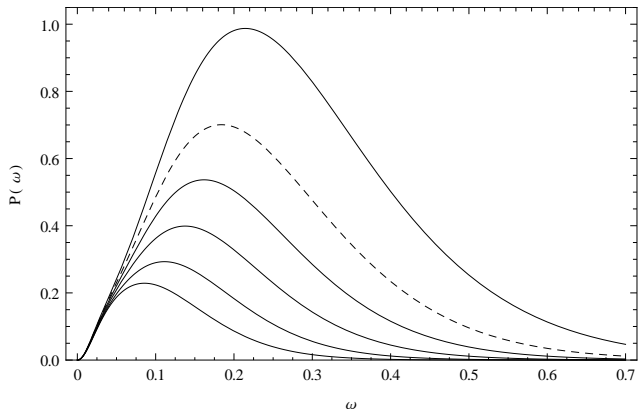}
  \label{fig:Powerspectrum}
  \caption{The frequency power spectrum of gravitational radiation emitted by an hyperbolic encounter. On the $x$-axis we have the angular frequency $\omega$ expressed in mHz units,
  whereas on the $y$-axis the amplitude of $P(\omega)$ is normalized to the maximum value of the $\varepsilon\sim2.5$ case.
  These are the expected emissions generated by a system of two supermassive black holes with $m = 10^7 M_{\odot}$, impact parameter $b=10$ AU, and different relative velocities.
  With lower velocities the interactions are stronger and the eccentricity decreases.
  These spectra, in order from the highest to the lowest, represent systems with $v_0 = 3.4 \times 10^7$ m/s ($\varepsilon \sim 2.5$), $v_0 = 3.5 \times 10^7$ m/s ($\varepsilon\sim3$),
  $v_0 = 3.6 \times 10^7$ m/s ($\varepsilon\sim3.1$), $v_0 = 3.75 \times 10^7$ m/s ($\varepsilon\sim3.4$), $v_0 = 4 \times 10^7$ m/s ($\varepsilon\sim3.8$),
  $v_0 = 4.5 \times 10^7$ m/s ($\varepsilon\sim4.7$), respectively.
  In particular the case with $\varepsilon\sim3$ (plotted with the dashed line) is discussed in the conclusions.
  As one can see, for higher eccentricities the peak frequency slowly decreases. This is only true for values of $v_0$ up to $\sim 6 \times 10^7$ m/s,
  whereas above it increases again. Moreover, decreasing the mass or increasing the impact parameter changes the eccentricity as well.
  We should be able to detect incoming waves in that range e.g. with eLISA, since the peak at $\sim 0.2$ mHz fits in its observable band.
  For a more detailed discussion see Sec. IV. and e.g. \cite{LISA}.}
 \end{center}
\end{figure}

We have checked the validity of this equality for different sets of values, comparable to those used in \cite{Capozziello},
e.g. $b=1\textmd{AU}$, $v_0 = 200$ km/s, and $m_{1,2}=1.4 \;\textmd{M}_{\odot}$, or similar.\\
For all of these sets we got agreement within numerical accuracy.

More interesting is the case where the eccentricity approaches $\varepsilon=1$.
According to eq. (\ref{eq:eccentricity}) this is the case e.g. with the set of initial conditions $b=2$ AU, $v_0 = 6.4$ km/s and $m_{1,2}=1.4\,M_{\odot}$.
Since this is a limit case for a parabolic trajectory, we can directly compare our result with the one studied by \cite{BerryGair},
and indeed they coincide, within numerical accuracy.
For a discussion about the feasibility of an analytical comparison see Appendix B.

\subsection{The limit for $\varepsilon \gg 1$}

As next we turn to the large  $\varepsilon$ limit and compare our result with the one given in \cite{Turner}
and \cite{WW}.
The expression for the total energy emitted during an hyperbolic interaction is written in \cite{Turner} as:
\begin{equation}\label{eq:E_Turner}
 \Delta E = \frac{8}{15} \frac{G^{7/2}}{c^5} \frac{m^{1/2}\,m_1^2\,m_2^2}{r_{min}^{7/2}} ~g(\varepsilon)~,
\end{equation}
where $r_{min}$ is the radius at periastron, and the enhancement factor $g(\varepsilon)$ turns out to be:
\begin{equation*}
 \frac{24 \arccos\big(\frac{-1}{\varepsilon}\big) \big(1+\frac{73}{24}\varepsilon^2+\frac{37}{96}\varepsilon^4\big)+\sqrt{\varepsilon^2-1}
 \big(\frac{301}{6}+\frac{673}{12}\varepsilon^2\big)}{(\varepsilon+1)^{7/2}}~.
\end{equation*}

Clearly this expression is equivalent to our result (\ref{eq:Delta_E}) for all values of $\varepsilon > 1$.
Indeed, using the following relations for the radius $r_{min}$ and the angle $\varphi_0$ at periastron:
\begin{equation*}
 \varepsilon = \frac{-1}{\cos\varphi_0}\,,\quad r_{min} = r(\varphi_0) = b\frac{\sin\varphi_0}{1-\cos\varphi_0}=\frac{Gm}{v_0^2}(\varepsilon-1)~,
\end{equation*}
both expressions can be written as functions of only three parameters describing the encounter: $\varepsilon$, $m$ and $v_0$, and it is then straightforward to see that
\begin{equation*}
 \Delta E_{Q} = \Delta E_{T}~,
\end{equation*}
where $\Delta E_{Q}$ denotes the energy computed according to eq. (\ref{eq:Delta_E}) and $\Delta E_{T}$ the one according 
to eq. (\ref{eq:E_Turner}).\\

Following \cite{Turner} again, we find out that in the limit for large $\varepsilon$, the $g(\varepsilon)$ factor can be simplified and written as:
\begin{equation}\label{eq:ge_semplice}
 \tilde{g}(\varepsilon) = \frac{37 \pi}{8}\sqrt{\varepsilon} \;+ \,\mathcal{O}(\,\varepsilon^{-1/2})~,
\end{equation}
which also agrees with the result of Wagoner and Will \cite{WW}.
This yields a simple form for the energy emitted during the path:
\begin{equation}\label{eq:E_Turner_simpl}
 \Delta \tilde{E}_T = \frac{8}{15} \frac{G^{7/2}}{c^5} \frac{m^{1/2}\,m_1^2\,m_2^2}{r_{min}^{7/2}} ~\tilde{g}(\varepsilon)~,
\end{equation}

That leads then to the formula for the energy spectrum valid for large $\varepsilon$:
\begin{equation}
\begin{split}
 P(\sigma) = \frac{G^{7/2}m^{1/2}\,m_1^2\,m_2^2}{c^5\quad r_{min}^{7/2}}\frac{8}{15\pi}\sqrt{\varepsilon}\;\times\;\bigg\{ 12\,[\sigma^2 K_2(\sigma)\quad\\
 - \, \sigma K_1(\sigma)]^2 + 3\,[\,2\sigma^2 K_1(\sigma)+\sigma K_0(\sigma)]^2 + \sigma^2 K_0^2(\sigma) \bigg\}~,
\end{split}
\end{equation}
where $K_{\tilde\alpha}(x)$ are the modified Bessel functions of the second kind, $\sigma$ is the frequency rescaled in terms of the characteristic time scale of the gravitational wave $\tau$
[=(periastron distance)/(periastron velocity)], $\sigma = \omega \tau$.\\

Comparing our total energy from the quadrupole approximation, eq. (\ref{eq:Delta_E}), with the expression for the energy $\Delta \tilde{E}_T$ (\ref{eq:E_Turner_simpl}) by \cite{Turner}
with the simplified factor (\ref{eq:ge_semplice}) valid in the large $\varepsilon$ limit, we see that they coincide for large eccentricities,
having e.g. a $1\%$ difference after $\varepsilon=100$, and a $5\%$ difference after $\varepsilon=20$ as shown in Fig. 4.

In fact we see that the behavior of the variation goes as $(\Delta E_{Q}-\Delta \tilde{E}_{T})/\Delta \tilde{E}_{T} \propto 1/\varepsilon$,
confirming the fact that for the parabolic limit $\Delta E_{Q} = 2 \, \Delta \tilde{E}_{T}$,
so that one would underestimate the energy emitted by a factor of $2$ taking this approximation \cite{Turner}.\\
\begin{figure}[!htb]
 \begin{center}
  \includegraphics[width=0.47\textwidth]{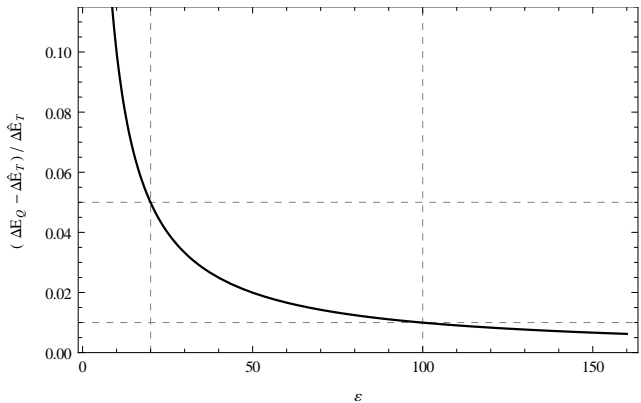}
  \label{fig:onepercent}
  \caption{The converging behavior of the total energy emitted in an hyperbolic encounter according to the quadrupole approximation $\Delta E_{Q}$,
  eq. (\ref{eq:Delta_E}), towards the result of \cite{Turner}, eq. (\ref{eq:E_Turner}).
  The plotted line is $(\Delta E_{Q}-\Delta \tilde{E}_{T})/\Delta \tilde{E}_{T}$ as function of the eccentricity $\varepsilon$.}
 \end{center}
\end{figure}

\section{Conclusions}\label{sec:conclusions}

Short gravitational wave burst-like signals are expected in the data stream of detectors.
Although these signals will likely be too short to allow us to measure the parameters of the emitting system accurately,
the results presented in this paper could be used to get a rough estimate of these parameters,
by observing the position of the peak, the amount of energy released and the timescale of the interaction.

Given the knowledge of the power spectrum we can easily see which kind of hyperbolic encounters could generate gravitational waves detectable e.g. with eLISA, advanced LIGO or advanced VIRGO.
Measurements from unbound interactions with ground-based detectors could in principle be possible, though the energy emitted at e.g. $\pm 200$ Hz
is below the minimum threshold for advanced LIGO or advanced VIRGO, making detections unlikely but not impossible.
The space-based interferometer instead is expected to cover frequencies ranging from $0.03$ mHz up to $1$ Hz (see e.g. \cite{LISA}), where the interactions could release more energy.

An unbounded collision between two intermediate-mass black holes, let's say of $10^3 M_{\odot}$ each, with an encounter velocity of $2000$ km/s at a distance of $1$ AU, would generate,
according to our eq. (\ref{eq:P_omega_mio}), a frequency spectrum with peak around $0.04$ mHz, with $80\%$ of the emission in the range between $0.01$ and $0.07$ mHz,
i.e. in the lower range limit of eLISA.
Another possible example of measurable impact would be an encounter between two supermassive black holes with mass, e.g., comparable to the expected mass of Sagittarius A*,
the black hole believed to be at the center of our galaxy, i.e. $\sim 10^7 M_{\odot}$.
With a distance of some AU, and a high velocity (we want to exclude the bounded case) of tens of thousands km/s, such a collision would generate
an energy spectrum with peak at $\sim 0.2$ mHz with $80\%$ between $0.03$ and $0.37$ mHz, thus in the observable range of eLISA. (Its energy spectrum is plotted with a dashed line in Fig. 3.)

Interestingly, the time window of such events is enough to allow measurements. Indeed, for encounters up to $\varepsilon \sim 5$ with peak in the mHz to Hz regime,
we are in the time scale of 1 day, if we choose the cutoff of the interaction at an angle of $\varphi = 3/4 \,\varphi_0$,
i.e. where the path starts to approach significantly the asymptote (see Fig. 1).
For instance, for the two examples above, the interactions would last about $54$h and $9$h, respectively.
Estimates for the rate of such events have been considered e.g. in \cite{cl}. They consider e.g. typical compact stellar cluster around the Galactic Center,
and expect an event rate of $10^{-3}$ up to unity per year, depending on the radius of the object and the amount of such clusters in the near region.

After a discussion with L. Blanchet we realized that there is a possibility to treat the problem in an alternative way.
Starting from the Keplerian equations of motion, one could take the solution of Peters and Mathews \cite{peters1},
bring the argument onto the imaginary axis and - making use of equations (9.6.2-9.6.4) in Abramowitz \& Stegun \cite{abr} - find the energy spectrum in terms of Hankel functions.

We believe that with the wave-form found here one should be able to classify the different encounters depending on the detected shape,
and therefore get a better insight into the map of our galaxy or the near universe.

\begin{acknowledgments}
We thank N. Straumann for useful discussions and for bringing to our attention the relevant treatment of the hyperbolic problem in electrodynamics in Landau \& Lifschitz.
We also thank L. Blanchet for his encouragement and for pointing out the possibility of treating the same problem in another way, as discussed in the conclusions.
Finally, we would also like to thank the referee for useful comments, and C. Berry for helping clarifying some details.
\end{acknowledgments}

\widetext

\vspace{0.005in}

\section*{APPENDIX\\~\\A. On the Fourier Transform of $\sinh\xi$ and $\cosh\xi$}
In section \ref{subsec:powerspec} we used the relation (\ref{eq:HankelRelations}) without any proof.
Since we didn't find any reference where the proof is shown explicitly, we will show in the following where these two relations arise from. In Landau and Lifshitz \cite{LL} the equations we used in the cited section are:
\begin{equation}\tag{A.1-A.2}
  \widehat{\sinh \xi} = - \frac{\pi}{\omega \varepsilon} H_{i \nu}^{(1)}(i \nu \varepsilon) \quad , \quad
  \widehat{\cosh \xi} = - \frac{\pi}{\omega} H_{i \nu}^{(1)}\textquotesingle(i \nu \varepsilon)
\end{equation}
but in fact the general relations we want to show - \cite{LL}, $\S$70, equation (70.15) - are:
\begin{equation}\tag{A.3-A.4}
  \widehat{y(t)} = \frac{a\,\sqrt{\varepsilon^2-1}\,\pi}{\omega \varepsilon} H_{i \nu}^{(1)}(i \nu \varepsilon) \quad , \quad
  \widehat{x(t)} = \frac{a\,\pi}{\omega} H_{i \nu}^{(1)\,\prime}(i \nu \varepsilon)
\end{equation}

\begin{proof}[\textbf{I. Fourier Transform of $y(t)$}]
We know that the time can also be parametrized with $\xi$ using the transformation:
\begin{equation*}
 t(\xi) = \sqrt{\frac{\mu\,a^3}{\alpha}}\,(\varepsilon\,\sinh\xi-\xi)\quad \textmd{which allows to write }\quad
 y(\xi) = a\,\sqrt{\varepsilon^2-1}\,\sinh\xi
\end{equation*}
We can now start with the computation of the Fourier transform:
\begin{align*}
 \widehat{y}(\omega) :&= \int_{-\infty}^{\infty}\!y(t)\,e^{-i\omega t}\,dt = \int_{-\infty}^{\infty}\!y(\xi)\,e^{-i \omega t(\xi)}\,\frac{dt(\xi)}{d\xi}\,d\xi
 = \int_{-\infty}^{\infty}\!y(\xi)\,\frac{-1}{i\omega}\frac{d}{d\xi}e^{-i \omega t(\xi)}\,d\xi =\\
 &= -\frac{a\,\sqrt{\varepsilon^2-1}}{i\,\omega}\int_{-\infty}^{\infty}\!\sinh\xi\,\frac{d}{d\xi}e^{-i \omega t(\xi)}\,d\xi
 \stackrel{(*^1)}= \frac{a\,\sqrt{\varepsilon^2-1}}{i\,\omega}\int_{-\infty}^{\infty}\!\cosh\xi\,e^{-i \omega t(\xi)}\,d\xi =\\
 &= \frac{a\,\sqrt{\varepsilon^2-1}}{i\,\omega}\int_{-\infty}^{\infty}\!(\cosh\xi-\frac{1}{\varepsilon}+\frac{1}{\varepsilon})\,e^{-i \omega t(\xi)}\,d\xi =\\
 &= \frac{a\,\sqrt{\varepsilon^2-1}}{i\,\omega\,\varepsilon}\,\bigg[\underbrace{\int_{-\infty}^{\infty}\!(\varepsilon\,\cosh\xi-1)\,e^{-i \omega t(\xi)}\,d\xi}_{\mathcal{A}} +
 \underbrace{\int_{-\infty}^{\infty}\!e^{-i \omega t(\xi)}\,d\xi}_{\mathcal{B}}\bigg]
\end{align*}
We now compute separately $\mathcal{A}$ and $\mathcal{B}$:
\begin{equation*}
 \mathcal{A} = \int_{-\infty}^{\infty}\!\frac{dt}{d\xi}\,\sqrt{\frac{\alpha}{\mu\,a^3}}\,e^{-i \omega t(\xi)}\,d\xi =
 \sqrt{\frac{\alpha}{\mu\,a^3}}\,\int_{-\infty}^{\infty}\!\frac{-1}{i\omega}\,\frac{d}{d\xi}\,e^{-i \omega t(\xi)}\,d\xi \stackrel{(*^2)}= 0
\end{equation*}
\begin{equation*}
 \mathcal{B} = \int_{-\infty}^{\infty}\!e^{-i\omega\sqrt{\frac{\mu a^3}{\alpha}}\,(\varepsilon\,\sinh\xi-\xi)}\,d\xi =
 \int_{-\infty}^{\infty}\!e^{-i\nu\varepsilon\,\sinh\xi\,+\,i\nu\xi}\,d\xi \stackrel{(*^3)}= i\pi\,H_{i\nu}^{(1)}(i\nu\varepsilon)
\end{equation*}
Hence we have:
\begin{equation}\tag{A.5}
 \widehat{y}(\omega) = \frac{a\,\sqrt{\varepsilon^2-1}}{i\,\omega\,\varepsilon}\,\bigg[0 + i\pi\,H_{i\nu}^{(1)}(i\nu\varepsilon) \bigg] =
 \frac{a\,\sqrt{\varepsilon^2-1}\,\pi}{\omega \varepsilon} H_{i \nu}^{(1)}(i \nu \varepsilon)
\end{equation}
\end{proof}

\noindent where: $(*^1)$ follows from partial integration and the boundaries vanish; we have the same situation for $(*^2)$;
and in $(*^3)$ we used the defining relation of the Hankel functions.
\newpage
\begin{proof}[\textbf{II. Fourier Transform of $x(t)$}]
We now show equation (A.4) in a similar way: again with the time parametrization through $\xi$ we can write:
\begin{equation*}
 x(\xi) = a\,(\varepsilon-\cosh\xi)
\end{equation*}
Its Fourier transform can be taken as follows:
\begin{align*}
 \widehat{x}(\omega) :&= \int_{-\infty}^{\infty}\!x(t)\,e^{-i\omega t}\,dt = \int_{-\infty}^{\infty}\!x(\xi)\,e^{-i\omega t(\xi)}\,\frac{dt(\xi)}{d\xi}\,d\xi =
 \int_{-\infty}^{\infty}\!x(\xi)\,\frac{-1}{i\omega}\,\frac{d}{d\xi}\,e^{-i\omega t(\xi)}\,d\xi =\\
 &= \frac{-a}{i\omega}\,\int_{-\infty}^{\infty}\!(\varepsilon-\cosh\xi)\,\frac{d}{d\xi}\,e^{-i\omega t(\xi)}\,d\xi = 
 \frac{-a}{i\omega}\,\int_{-\infty}^{\infty}\!\sinh\xi\,e^{-i\omega t(\xi)}\,d\xi =\\
 &= \frac{-a}{i\omega}\,\int_{-\infty}^{\infty}\!\frac{e^\xi-e^{-\xi}}{2}\,e^{-i\omega t(\xi)}\,d\xi =
 \frac{-a}{i\omega}\,\int_{-\infty}^{\infty}\!\frac{e^\xi-e^{-\xi}}{2}\,e^{-i\nu\varepsilon\sinh\xi+i\nu\xi}\,d\xi =\\
 &=\frac{-a}{i\omega}\,\frac{1}{2}\,\bigg[\int_{-\infty}^{\infty}\!e^\xi\,e^{-i\nu\varepsilon\sinh\xi+i\nu\xi}\,d\xi -
 \int_{-\infty}^{\infty}\!e^{-\xi}\,e^{-i\nu\varepsilon\sinh\xi+i\nu\xi}\,d\xi\; \bigg] =\\
 &=\frac{-a}{i\omega}\,\frac{1}{2}\,\bigg[\int_{-\infty}^{\infty}\!e^{-i\nu\varepsilon\sinh\xi+(i\nu+1)\xi}\,d\xi -
 \int_{-\infty}^{\infty}\!e^{-i\nu\varepsilon\sinh\xi+(i\nu-1)\xi}\,d\xi\; \bigg] =\\
 &=\frac{-a}{i\omega}\,\frac{1}{2}\,\bigg[ i\pi\,H^{(1)}_{i\nu+1}(i\nu\varepsilon) - i\pi\,H^{(1)}_{i\nu-1}(i\nu\varepsilon) \bigg] =
 \frac{a\,\pi}{\omega}\,\frac{1}{2}\,\bigg[ H^{(1)}_{i\nu-1}(i\nu\varepsilon) - H^{(1)}_{i\nu+1}(i\nu\varepsilon) \bigg]
\end{align*}
Hence, using the formula for the derivative given in (\ref{eq:HankelDerivative}) we get:
\begin{equation}\tag{A.6}
 \widehat{x}(\omega) = \frac{a\,\pi}{\omega} H_{i \nu}^{(1)\,\prime}(i \nu \varepsilon)
\end{equation}
\end{proof}

\section*{B. Analytical comparison of the parabolic limit}
In section \ref{subsec:powerspec} we discuss the parabolic limit of our result for the energy spectrum, where $\varepsilon \rightarrow 1$.
A numerical check of its validity with respect to the previous known quantity, in \cite{BerryGair}, is quite straightforward.
Nevertheless it would be interesting to find an analytical agreement between these two formulae.\\
Taking that limit the last part of our equations drops away, and only three terms remain.
These can be greatly simplified, since in the $\varepsilon = 1$ limit the second and third are similar and $F_{\varepsilon}(\omega)$ becomes:
\begin{equation*}\tag{B.1}
 F_{\varepsilon}(\omega) = [\,16\;H^{(1)\prime}_{i\nu}(i\nu\varepsilon)\,-2\;\,H^{(1)\prime}_{i\nu}(i\nu\varepsilon/2)\,]^2 + 128 \;[\,H^{(1)\prime}_{i\nu}(i\nu\varepsilon)\,]^2 + 2\;[\,H^{(1)\prime}_{i\nu}(i\nu\varepsilon/2)\,]^2
\end{equation*}
Following the book by Abramowitz \& Stegun \cite{abr} and the work done in \cite{BerryGair} we see two facts.
First, the frequency in the order of these functions goes as $(1-\varepsilon)^{-3/2}$ (see eq. (10) in \cite{BerryGair}), and therefore in the limit $\varepsilon\rightarrow1$, $\nu$ goes to infinity.
Second, we note that the order is also in the argument, therefore when $\nu \rightarrow \infty$ we can write the Hankel functions in terms of Airy functions $\mathrm{Ai}(x)$, e.g. eq. (9.3.45) in \cite{abr}:
\begin{equation*}\tag{B.2}
 H^{(1)\prime}_{i\nu}(i\nu\varepsilon)\; \sim\; \textmd{coeff.}\, \times \bigg\{ \frac{\mathrm{Ai}(e^{2\pi i/3}\nu^{2/3}\zeta)}{\nu^{4/3}} \sum_{k=0}^{\infty}\frac{c_k(\zeta)}{\nu^{2k}} +
 \frac{\mathrm{Ai}^{\prime}(e^{2\pi i/3}\nu^{2/3}\zeta)\,e^{2\pi i/3}}{\nu^{2/3}} \sum_{k=0}^{\infty}\frac{d_k(\zeta)}{\nu^{2k}} \bigg\}
\end{equation*}
where $\zeta$ is a function of $\varepsilon$ defined as $\frac{2}{3}\zeta^{3/2}=\ln \frac{1+\sqrt{1-\varepsilon^2}}{\varepsilon}-\sqrt{1-\varepsilon^2}$, see eq. (9.3.38) in \cite{abr}.
If we only take the first order of the series, we have for the derivatives a term in $\mathrm{Ai}(x)$ and one in $\mathrm{Ai}^{\prime}(x)$.
Again using \cite{abr} we find eq. (10.4.26 - 10.4.31) which express these terms in form of modified Bessel functions of the second kind:
\begin{equation*}\tag{B.3}
 K_{\pm 1/3}(\zeta) = \pi \sqrt{3/z}\,\mathrm{Ai}(z) \quad,\quad K_{\pm 2/3}(\zeta) = -\pi \sqrt{3}/z\,\mathrm{Ai}^{\prime}(z)
\end{equation*}
where $z = (\frac{3}{2}\zeta)^{2/3}$. We see now that equation (B.1) has the very same structure as eq. (24) in \cite{BerryGair}:
\begin{equation*}\tag{B.4}
 \ell(\tilde{f}) = \big[\,8 \mathcal{B}(\tilde{f}) - 2 \mathcal{A}(\tilde{f})\,\big]^2 + C \;[\,\mathcal{A}(\tilde{f})\,\big]^2 + D \; [\,\mathcal{A}(\tilde{f})\,\big]^2
\end{equation*}
where $\mathcal{A}(\tilde{f})$ and $\mathcal{B}(\tilde{f})$ are also in terms of modified Bessel functions of the second kind of order $\pm1/3$, $\pm2/3$.\\
At this point with some cumbersome algebra one should find that also all the coefficients of (B.2) and (B.4) agree.

\end{document}